\documentclass[final,5p,times,twocolumn,numbers]{elsarticle}

\usepackage{amssymb}
\usepackage{lipsum}
\usepackage{amsmath}
\usepackage{graphicx}
\usepackage{caption}
\usepackage{subcaption}
\usepackage{float}
\usepackage{color}
\usepackage{booktabs}

\journal{Nuclear Physics B}

\begin{document}

\begin{frontmatter}

\title{Exploring Pion-Induced High-Momentum Components in Nuclei via $(p,p'\pi)$ Reactions}

\author[first,second]{Junki Tanaka\corref{cor1}}\ead{junki@rcnp.osaka-u.ac.jp}\cortext[cor1]{Corresponding author}
\author[second]{Junichi Kato}
\author[first]{Hiroshi Toki}

\affiliation[first]{organization={Research Center for Nuclear Physics, The University of Osaka},
            addressline={10-1 Mihogaoka},
            city={Ibaraki},
            postcode={567-0047},
            state={Osaka},
            country={Japan}}

            \affiliation[second]{organization={Department of Physics, The University of Osaka},
            addressline={1-1 Machikaneyama-cho},
            city={Toyonaka},
            postcode={560-0043},
            state={Osaka},
            country={Japan}}

\begin{abstract}
Pion exchange plays a fundamental role in nuclear structure and is responsible for tensor correlations and high-momentum components in nuclei. The $(p,p'\pi)$ reaction provides a unique opportunity to investigate pion dynamics under large-momentum-transfer conditions. Its three-body kinematics allows large momentum transfer to be achieved while keeping the excitation energy of the residual nucleus low.
We investigate the kinematical properties of the $^{12}\mathrm{C}(p,p'\pi^+)^{12}\mathrm{B}$ reaction using Lorentz-invariant three-body phase-space calculations. The calculations were performed for a 392-MeV proton beam assuming a constant transition amplitude. The resulting momentum-transfer map and phase-space distribution identify experimentally accessible regions of large momentum transfer and provide guidance for optimizing a double-arm spectrometer experiment at RCNP.
The present study establishes a model-independent kinematical foundation for future investigations of pion-induced correlations, high-momentum components, and pion dynamics in nuclei.
\end{abstract}


\end{frontmatter}




\section{Introduction}
\vspace{-5pt}
Pion exchange is a fundamental component of the nuclear force and plays a central role in generating tensor correlations and high-momentum components in nuclei\cite{EricsonWeise1988}. Although its importance has long been recognized, experimental access to pion degrees of freedom in nuclei remains limited.

The $(p,p'\pi)$ reaction provides a unique opportunity to investigate pion dynamics in nuclei. Because the final state contains an emitted pion, the reaction is sensitive to spin-isospin excitations associated with pion dynamics in nuclei. Moreover, its three-body kinematics allows a substantial fraction of the transferred energy and momentum to be carried away by the pion. As a result, large momentum transfer can be achieved while keeping the residual nucleus at low excitation energy, a condition difficult to realize in conventional two-body reactions.

In this work, we investigate the $^{12}$C$(p,p'\pi^+)^{12}$B reaction at an incident proton energy of 392 MeV. Lorentz-invariant three-body phase-space calculations are performed to examine the kinematical accessibility of low-lying final states under large momentum-transfer conditions. The calculations provide a model-independent baseline for future studies of pion-induced correlations, high-momentum components, and pion dynamics in nuclei.

\section{Pion Dynamics and High-Momentum Components}
\vspace{-5pt}
The pion-exchange interaction plays a fundamental role in nuclear structure and provides one of the primary mechanisms responsible for high-momentum components in nuclear wave functions. Its momentum-space structure can be expressed as
\begin{equation}
\frac{(\boldsymbol{\sigma}_1\cdot\mathbf q)(\boldsymbol{\sigma}_2\cdot\mathbf q)}{m_\pi^2+q^2}
=\frac{1}{3}(\boldsymbol{\sigma}_1\cdot\boldsymbol{\sigma}_2)
\left(1-\frac{m_\pi^2}{m_\pi^2+q^2}\right)
+\frac{1}{3}S_{12}(\hat{\mathbf q})
\frac{q^2}{m_\pi^2+q^2},
\label{eq:pion_tensor}
\end{equation}
where $S_{12}$ denotes the tensor operator. The second term represents the tensor component of the pion-exchange interaction. Because it is proportional to $q^2/(m_\pi^2+q^2)$, pion exchange naturally generates short-range and high-momentum components in nuclear wave functions. In nuclei, the tensor interaction couples configurations differing by two units of orbital angular momentum ($\Delta L=2$), producing two-particle--two-hole excitations beyond the mean-field picture. Modern calculations, including the Tensor-Optimized Shell Model (TOSM)\cite{Myo2008}, have demonstrated that such pion-induced tensor correlations contribute significantly to nuclear binding and high-momentum components.

Pion exchange is one of the primary mechanisms generating high-momentum components in nuclei. In the nuclear medium, pion propagation is modified through coupling to particle-hole and $\Delta$-hole excitations. Such medium effects can give rise to collective spin-isospin excitations involving explicit pion degrees of freedom, often referred to as pionic modes in nuclei \cite{Toki1979}. Understanding how pion dynamics manifest themselves in finite nuclei therefore remains an important problem in nuclear many-body physics.

Because the tensor component of the pion-exchange interaction increases with momentum transfer, experimental investigations of pion-induced correlations require access to the large-momentum-transfer region. Conventional two-body reactions generally couple large momentum transfer to large excitation energy, making it difficult to study pion-related phenomena in low-lying nuclear states. As discussed in the following section, the three-body kinematics of the $(p,p'\pi)$ reaction provides a unique opportunity to overcome this limitation by accessing large momentum transfer while maintaining relatively low excitation energy in the residual nucleus.

\section{Kinematical Advantage of the $(p,p'\pi)$ Reaction}
\vspace{-5pt}
A central advantage of the $(p,p'\pi)$ reaction is that the excitation energy of the residual nucleus and the momentum transfer can be controlled almost independently. In ordinary two-body reactions,
\begin{equation}
a+A \rightarrow b+B^{*},
\end{equation}
energy and momentum conservation establish an almost one-to-one correspondence between the excitation energy $E_x$ and the momentum transfer $q$. Consequently, large momentum transfer is generally accompanied by large excitation energy.

In contrast, the $(p,p'\pi)$ reaction has a three-body final state,
\begin{equation}
p+A \rightarrow p'+\pi+B,
\end{equation}
for which energy and momentum conservation are given by
\begin{align}
E_p+M_A &= E_{p'}+E_\pi+E_B, \\
\mathbf{p}_p &= \mathbf{p}_{p'}+\mathbf{p}_{\pi}+\mathbf{P}_B.
\end{align}
The momentum transfer to the residual nucleus is therefore
\begin{equation}
\mathbf{q}=\mathbf{P}_B=\mathbf{p}_p-\mathbf{p}_{p'}-\mathbf{p}_{\pi}.
\label{eq:qtransfer}
\end{equation}
Because the emitted pion can carry a substantial fraction of the transferred energy and momentum, large momentum transfer can be achieved while keeping the residual nucleus in its ground state or low-lying excited states. For the reaction
\begin{equation}
{}^{12}\mathrm{C}(p,p'\pi^+){}^{12}\mathrm{B},
\end{equation}
the residual nucleus remains bound, allowing pion-induced correlations to be investigated in low-lying nuclear states under large-momentum-transfer conditions.

A particularly important feature of the present experimental configuration is that the scattered proton is measured at forward angles close to $0^\circ$. In this geometry, the momentum transfer can be determined experimentally from the measured proton and pion momenta.
Within a pion-exchange picture, the momentum transfer reflects the momentum scale associated with pion exchange inside the nucleus. Large momentum transfer is expected to enhance sensitivity to pion-induced correlations and related high-momentum components. Consequently, measurements at large momentum transfer are expected to provide enhanced sensitivity to pion dynamics in nuclei.

\begin{figure}[t]
\centering
\includegraphics[width=0.95\columnwidth]{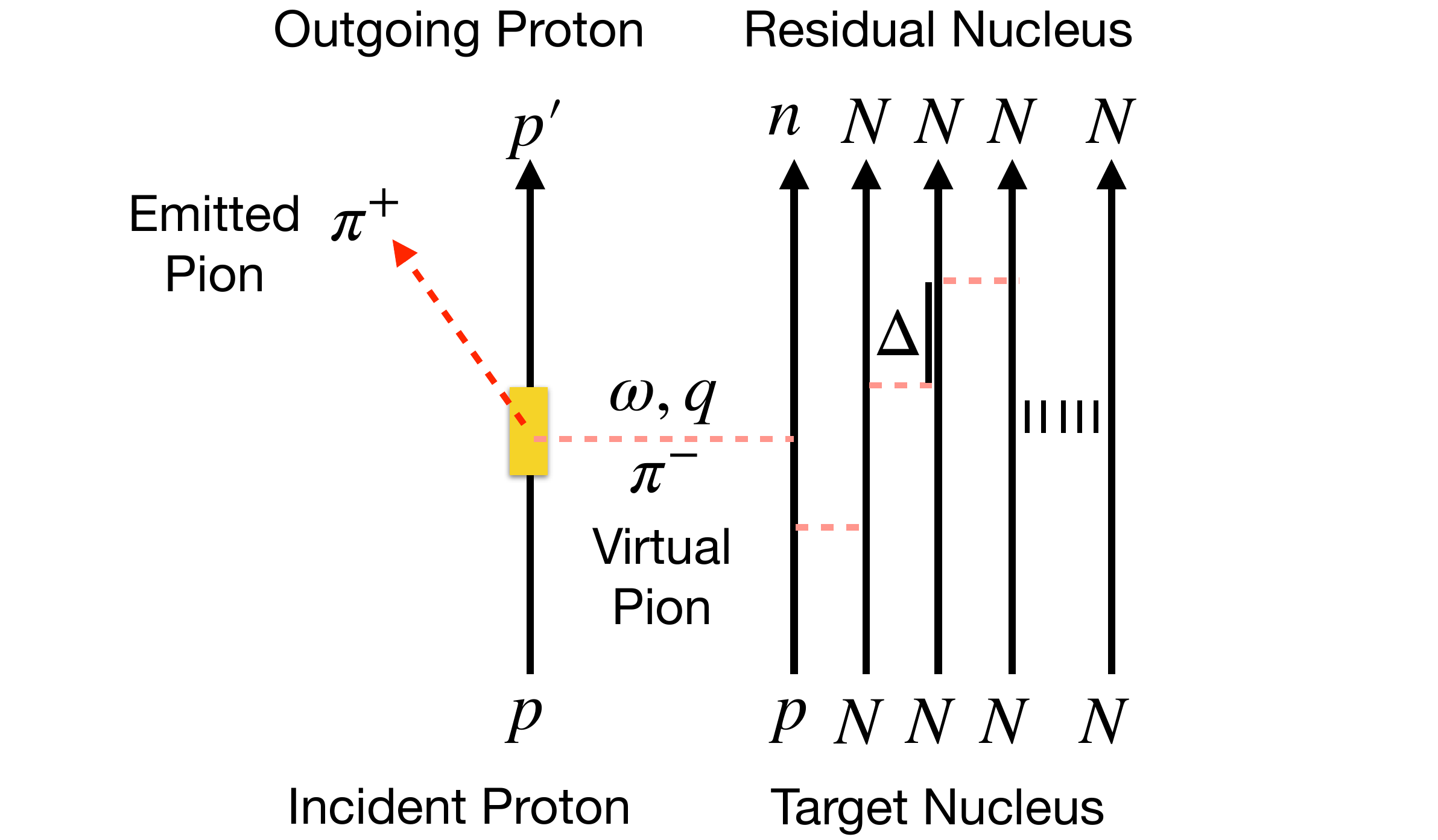}
\caption{
Schematic illustration of pion exchange in the $(p,p'\pi)$ reaction. The exchanged virtual pion is schematically associated with the momentum transfer to the nucleus. Because the emitted pion provides an additional degree of freedom in the three-body final state, large momentum transfer can be achieved while maintaining the residual nucleus in low-lying states.
}
\label{fig:concept}
\end{figure}

Figure~\ref{fig:concept} illustrates this physical picture schematically. The incident proton interacts with the nucleus through pion exchange, leading to pion emission and a recoiling residual nucleus. By measuring the scattered proton and emitted pion, the momentum transfer can be determined event by event.
The emitted pion provides an additional degree of freedom that is absent in ordinary two-body reactions. As a result, the $(p,p'\pi)$ reaction can access large momentum-transfer conditions while keeping the residual nucleus in its ground state or low-lying excited states. This feature enables state-by-state spectroscopy under large-momentum-transfer conditions and provides a favorable experimental environment for investigating pion-induced correlations and high-momentum components in nuclei.

\section{Three-Body Kinematics and Phase-Space Calculation}
\begin{figure}[t]
\centering
\includegraphics[width=1.0\columnwidth]{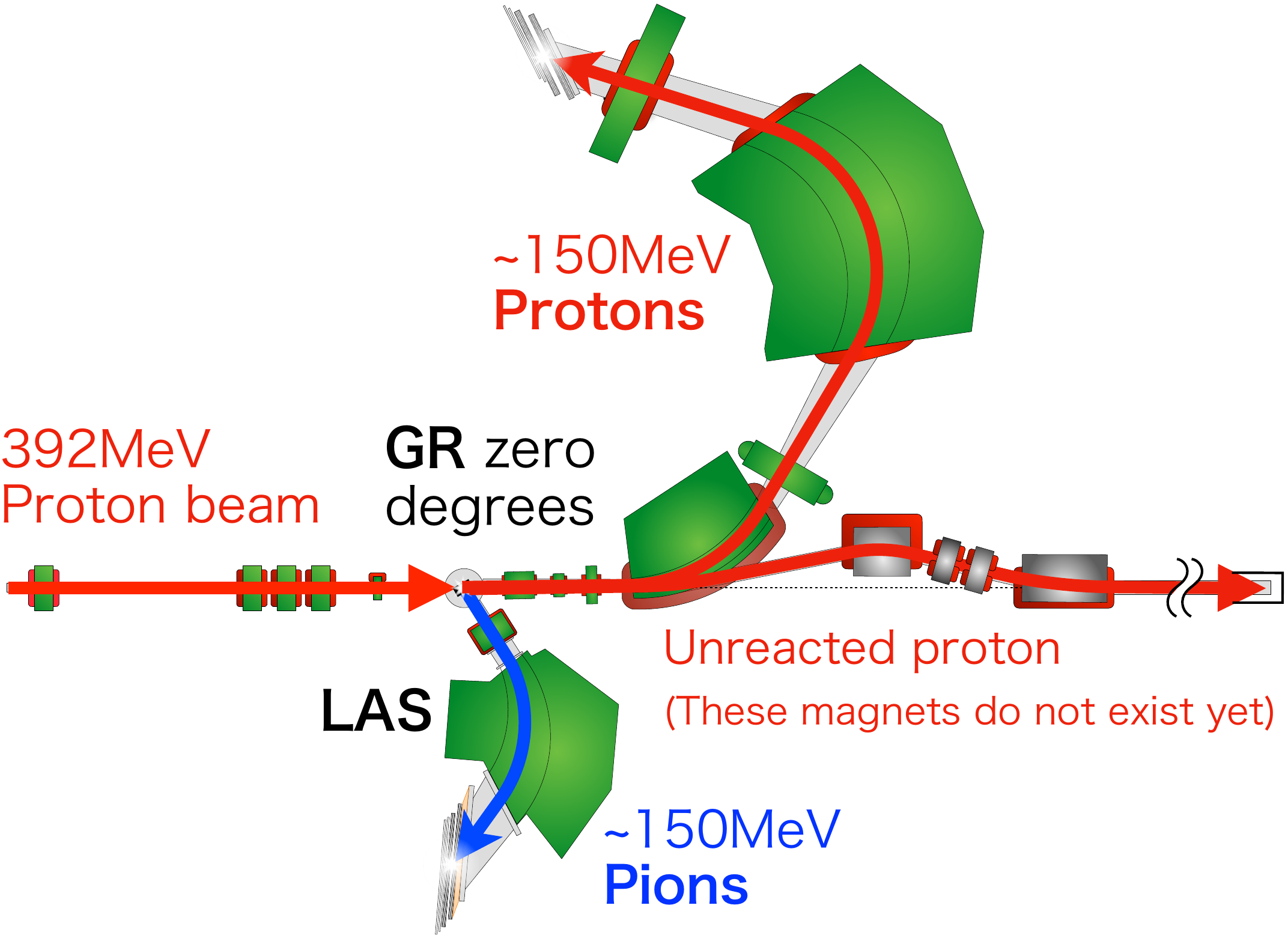}
\caption{
Assumed experimental configuration for the
$^{12}\mathrm{C}(p,p'\pi^+)^{12}\mathrm{B}$
reaction using the GR--LAS spectrometer system at RCNP.
}
\label{fig:setup}
\end{figure}

\vspace{-5pt}
To investigate the kinematical accessibility of the large-momentum-transfer region discussed in the previous section, we perform Lorentz-invariant three-body phase-space calculations for the $^{12}\mathrm{C}(p,p'\pi^+)^{12}\mathrm{B}$ reaction.
The calculation assumes an incident proton energy of 392 MeV and a reaction $Q$ value of $-153$ MeV. The scattered proton is assumed to be measured by the Grand Raiden (GR) spectrometer \cite{Fujiwara1999} at forward angles close to $0^\circ$, while charged pions are detected by the Large Acceptance Spectrometer (LAS) \cite{Matsuoka1995, Miyagawa2026}. A schematic layout of the assumed experimental configuration is shown in Fig.~\ref{fig:setup}.
The reaction cross section is expressed as
\begin{equation}
d\sigma=\frac{1}{v}|\mathcal{M}|^2 d\Phi,
\end{equation}
where $v$ is the relative velocity in the initial state, $\mathcal{M}$ is the reaction amplitude, and $d\Phi$ denotes the Lorentz-invariant three-body phase-space element. In the present exploratory study, Fermi motion, initial- and final-state interactions, and nuclear transition amplitudes are neglected. Furthermore, the transition amplitude is assumed to be constant. Under these assumptions, the event distribution is determined entirely by the available phase-space volume.
The Lorentz-invariant phase-space element is given by
\begin{equation}
d\Phi=(2\pi)^4\delta^{(4)}(P-p'-k-P_B)
\frac{d^3p'}{(2\pi)^3 2E_{p'}}
\frac{d^3k}{(2\pi)^3 2E_{\pi}}
\frac{d^3P_B}{(2\pi)^3 2E_B},
\end{equation}
where $P=p+P_A$ denotes the total four-momentum of the initial system. For numerical calculations, the phase space is factorized into two sequential two-body phase spaces,
\begin{equation}
d\Phi(P;p',k,P_B)=
\frac{dM_{p\pi}^2}{2\pi}
\Phi_2(P;M_{p\pi},M_B)
\Phi_2(M_{p\pi};m_p,m_\pi),
\end{equation}
where
\begin{equation}
\Phi_2(s;m_1,m_2)=
\frac{1}{8\pi}
\frac{\sqrt{\lambda(s,m_1^2,m_2^2)}}{s},
\end{equation}
with the K{\"a}ll{\'e}n function
\begin{equation}
\lambda(x,y,z)=x^2+y^2+z^2-2xy-2yz-2zx.
\end{equation}
Events were generated according to the Lorentz-invariant phase-space distribution using Monte Carlo sampling. The invariant mass of the intermediate $(p'\pi)$ subsystem was sampled according to the corresponding phase-space weight, and the final-state momenta were generated isotropically in the relevant rest frames before being transformed to the laboratory frame through successive Lorentz boosts.

\begin{figure}[t]
\centering
\includegraphics[width=1.1\columnwidth]{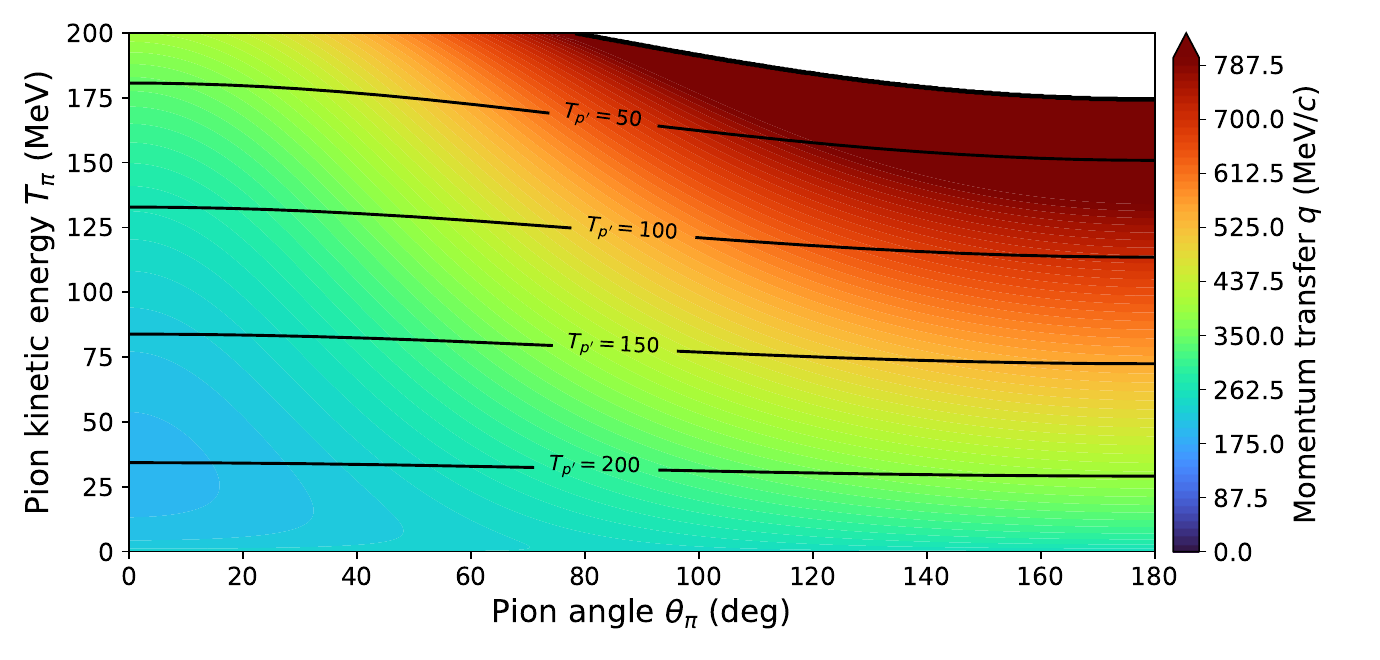}
\includegraphics[width=1.1\columnwidth]{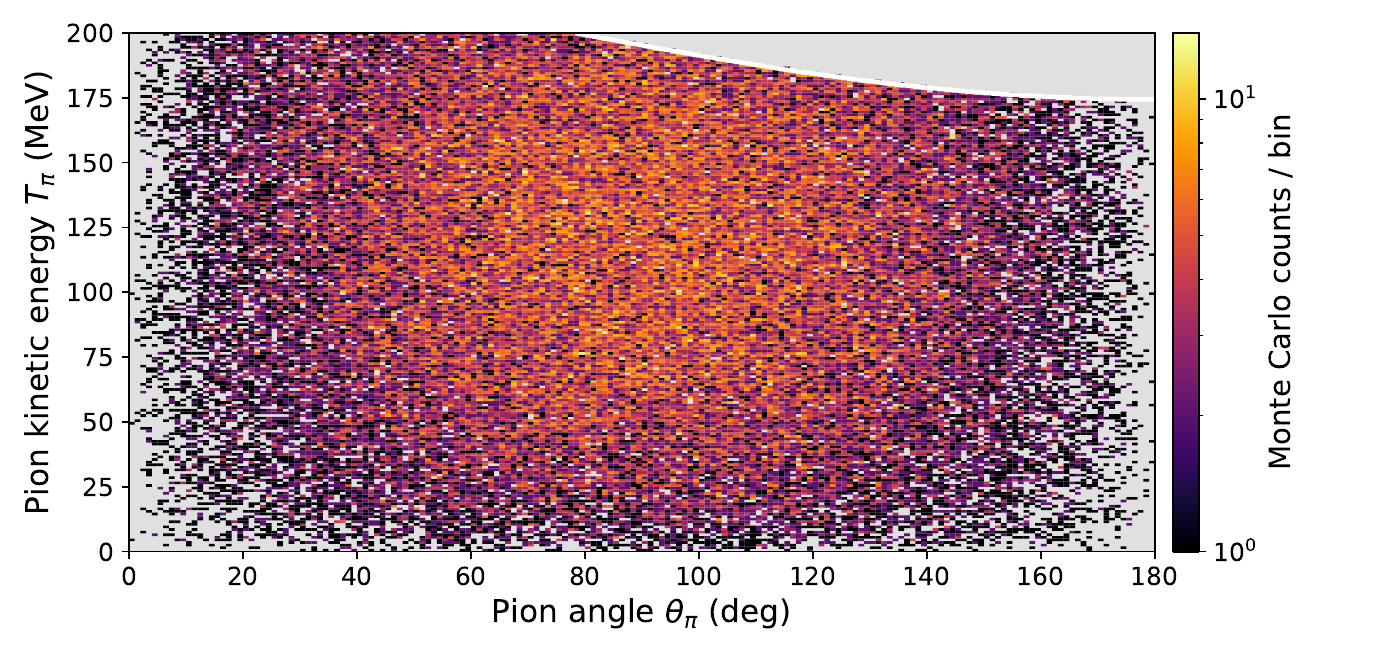}
\caption{
(a) Momentum transfer for the
$^{12}\mathrm{C}(p,p'\pi^+)^{12}\mathrm{B}(\mathrm{g.s.})$
reaction.
The color scale represents the magnitude of the momentum transfer. The corresponding scattered-proton kinetic energies are shown in parentheses. The black curve indicates the kinematical boundary for the ground-state transition.
(b) Lorentz-invariant three-body phase-space density projected onto the $(\theta_\pi,T_\pi)$ plane. Events were generated assuming a constant transition amplitude and requiring $\theta_{p'}<10^\circ$. The white curve indicates the kinematical boundary for the ground-state transition.
}
\label{fig:kinematics}
\end{figure}

Figure~\ref{fig:kinematics}(a) shows the momentum transfer for the ground-state transition $^{12}\mathrm{B}(\mathrm{g.s.})$. The color scale represents the magnitude of the momentum transfer, while the contours indicate constant scattered-proton kinetic energies. The figure demonstrates that large momentum transfer is preferentially accessed in the region of large pion kinetic energies and large pion emission angles. In particular, the upper-right region corresponds to the largest momentum transfer and is therefore expected to provide enhanced sensitivity to pion-induced correlations and high-momentum components in nuclei.
Kinematical accessibility alone, however, does not guarantee experimentally useful event yields. Figure~\ref{fig:kinematics}(b) therefore shows the Lorentz-invariant three-body phase-space density projected onto the $(\theta_\pi,T_\pi)$ plane. Events were generated assuming a constant transition amplitude and only events satisfying $\theta_{p'}<10^\circ$ were retained. 
The phase-space distribution extends over a broad fraction of the kinematically allowed region. Although the event density gradually decreases toward the extreme large-angle and large-energy region, the phase-space suppression remains moderate. Consequently, the large-momentum-transfer region remains experimentally accessible and is not excluded by phase-space limitations alone.
Taken together, Figs.~\ref{fig:kinematics}(a) and \ref{fig:kinematics}(b) demonstrate that the $(p,p'\pi)$ reaction provides access to large momentum transfer under experimentally accessible conditions. This unique combination makes the reaction a promising tool for future investigations of pion-induced correlations, high-momentum components, and pion dynamics in nuclei.

\section{Conclusion and Outlook}
\vspace{-5pt}

We have investigated the kinematical properties of the $^{12}\mathrm{C}(p,p'\pi^+)^{12}\mathrm{B}$ reaction at an incident proton energy of 392 MeV using Lorentz-invariant three-body phase-space calculations.
The results demonstrate that the three-body kinematics of the $(p,p'\pi)$ reaction allows access to large momentum transfer while keeping the residual nucleus in low-lying states.The corresponding phase-space calculation indicates that this region remains experimentally accessible. These characteristics make the $(p,p'\pi)$ reaction a promising tool for investigating pion-induced correlations and high-momentum components in nuclei.

The present calculation neglects Fermi motion, initial- and final-state interactions, and nuclear transition amplitudes. Future studies will incorporate these effects together with realistic nuclear wave functions and microscopic reaction amplitudes including pion exchange and $\Delta$-hole excitations. The present approach can also be extended to higher beam energies and larger momentum transfers, allowing studies of both low-lying states and higher excitation-energy regions.

\section*{Acknowledgment}

This work was supported by the Yamada Science Foundation through a Research Grant in 2025.


\begin{thebibliography}{9}
\bibitem{EricsonWeise1988}
T.~E.~O.~Ericson and W.~Weise,
\textit{Pions and Nuclei},
Oxford University Press, Oxford (1988).
\bibitem{Myo2008}
T.~Myo, K.~Kato, H.~Toki, and K.~Ikeda,
Prog. Theor. Phys. \textbf{119}, 561 (2008).
\bibitem{Toki1979}
H.~Toki and W.~Weise,
Phys. Rev. Lett. \textbf{42}, 1034 (1979).
\bibitem{Fujiwara1999}
M.~Fujiwara, H.~Akimune, I.~Daito, H.~Ejiri, K.~Fujita,
Y.~Fujita, M.~N.~Harakeh, T.~Hashimoto, K.~Hatanaka,
M.~Hosaka, S.~Kawakami, H.~Kuboki, Y.~Maeda, H.~Miyatake,
M.~Nakamura, T.~Noro, A.~Sakai, Y.~Shimizu, Y.~Tonegawa,
H.~Toyokawa, M.~Yosoi,
Nucl. Instrum. Methods Phys. Res. A \textbf{422}, 484 (1999).
\bibitem{Matsuoka1995}
N.~Matsuoka, T.~Noro, H.~Sakaguchi, T.~Yabe,
Nucl. Instrum. Methods Phys. Res. A \textbf{345}, 1 (1994).
\bibitem{Miyagawa2026}
T.~Miyagawa, J.~Tanaka, S.~Ota, M.~Dozono,
N.~Kobayashi, L.~Wickremasinghe, F.~Furukawa,
D.~Ishii, S.~Nishioka, K.~Takahashi, and E.~Ukai,
arXiv:2605.08127 (2026).
\end{thebibliography}
\end{document}